\documentclass[twocolumn,superscriptaddress,amsmath,amssymb]{revtex4}
\usepackage{graphicx}
\usepackage{dcolumn}
\usepackage{bm}
\usepackage{color}

\usepackage[colorlinks,
           bookmarks=true,
           linkcolor=blue,
           urlcolor=blue,
            anchorcolor=black,
            citecolor=blue
            ]{hyperref}

\usepackage[all]{hypcap}

\begin{document}

\title{Machine learning phase transitions of the three-dimensional Ising universality class}

\author{Xiaobing Li}
\affiliation{Key Laboratory of Quark and Lepton Physics (MOE) and Institute of Particle Physics, \\
Central China Normal University, Wuhan 430079, China}
\author{Ranran Guo}
\affiliation{Key Laboratory of Quark and Lepton Physics (MOE) and Institute of Particle Physics, \\
Central China Normal University, Wuhan 430079, China}
\author{Yu Zhou}
\affiliation{University of California, Los Angeles, CA 90095, USA}
\author{Kangning Liu}
\affiliation{Key Laboratory of Quark and Lepton Physics (MOE) and Institute of Particle Physics, \\
Central China Normal University, Wuhan 430079, China}
\author{Jia Zhao}
\affiliation{Key Laboratory of Quark and Lepton Physics (MOE) and Institute of Particle Physics, \\
Central China Normal University, Wuhan 430079, China}
\author{Fen Long}
\affiliation{Key Laboratory of Quark and Lepton Physics (MOE) and Institute of Particle Physics, \\
Central China Normal University, Wuhan 430079, China}
\author{Yuanfang Wu}
\affiliation{Key Laboratory of Quark and Lepton Physics (MOE) and Institute of Particle Physics, \\
Central China Normal University, Wuhan 430079, China}
\author{Zhiming Li}
\email{lizm@mail.ccnu.edu.cn}
\affiliation{Key Laboratory of Quark and Lepton Physics (MOE) and Institute of Particle Physics, \\
Central China Normal University, Wuhan 430079, China}

\begin{abstract}
Exploration of the QCD phase diagram and critical point is one of the main goals in current relativistic heavy-ion collisions. The QCD critical point is expected to belong to a three-dimensional (3D) Ising universality class. 
Machine learning techniques are found to be powerful in distinguishing different phases of matter and provide a new way to study the phase diagram. We investigate phase transitions in the 3D cubic Ising model  using supervised learning methods. It is found that a 3D convolutional neural network can be trained to effectivelly predict physical quantities in different spin configurations. With a uniform neural network architecture, it can encode phases of matter and identify both  second- and first-order phase transitions. The important features that discriminate different phases in the classification processes are investigated. These findings can help  study and understand QCD phase transitions in relativistic heavy-ion collisions.
\end{abstract}

\maketitle
\section{Introduction}
One of the main goals in the study of high energy heavy-ion collisions is to explore the QCD phase structure and critical point (CP)~
\cite{StephanovPD,adams2005experimental,conservecharge0,conservecharge1}. Owing to the fermion sign problem, Lattice QCD calculation is restricted to the region of a vanishing or small baryon chemical potential ($\mu_B$) and predicts a crossover from hadronic phase to a quark gluon plasma (QGP) phase in this area~\cite{Lattice, Crossover}. The results of QCD based models indicate that the transition could be first-order at large $\mu_B$~\cite{firstorder}. The point where the first-order phase transition ends is the CP~\cite{CEP1, CEP2}. This CP is proposed to be characterized by a second-order phase transition, which becomes a unique property of strongly interacting matter~\cite{searchCEP1,searchCEP2,searchCEP3,searchCEP4}. Attempts are being made to explore the CP and phase boundary both experimentally and theoretically~\cite{searchCEP1,searchCEP2,searchCEP3,searchCEP4,net_proton2010,net_proton2014,net_charge2014,net-kaon,phenix,luo2015energy,ExpeReview,TheoReview}.

The QCD equation of state with a CP is an essential ingredient for hydrodynamic simulations of fireball evolution in heavy-ion collision. The universality of critical phenomena allows us to predict the leading singularity, $i.e.$, the leading term of the Taylor expansion, of the QCD equation of state near the CP~\cite{TheoReview}. Systems with the same symmetry in the same dimension share the same critical behaviour, even though they are governed by different interactions. It is argued that the QCD CP belongs to the same Z(2) universality class~\cite{QCDZ21,QCDZ22} as the three-dimensional (3D) Ising model~\cite{StephanovPD,Mapping1,Mapping2,Mapping3,Mapping4}. Therefore, universality makes the Ising model very relevant for studies of systems that display the Z(2) symmetry~\cite{IsingZ21,IsingZ22,IsingZ23,IsingZ24}. 
Via parameterization of the scaling equation of state in the 3D Ising model and non-universal mapping, it allows us to construct an equation of state matching the first principle lattice QCD calculations and include the proper scaling behavior in the proximity of the CP~\cite{searchCEP2,mapEOS1,mapEOS2,mapEOS3,mapEOS4}. It can also map the phase diagram of the 3D Ising model onto the one of QCD~\cite{mapEOS1,mapPhaseD1,mapPhaseD2}. Thus, the CP, the lines of the first-order phase transition, and crossover in the 3D Ising model are related to those of QCD. 

Classifying phases of matter and identifying phase transitions are central topics in current phase structure investigations. In the conventional statistical method, these rely on the identification of order parameters or the analysis of singularities in the free energy and its derivatives. However, the order parameter of the QCD phase transition is difficult to determine and measure in experiments. Moreover, certain phases, such as topological ones~\cite{topology1,topology2}, do not have any clear order parameters  because the symmetry that drives the phase transition is not manifest in the Hamiltonian. Therefore, it is important to develop new methods of identifying phases in these systems. With the development of more powerful computers and artificial neural networks, machine learning (ML)~\cite{ML1,ML2}, a data-driven method, is proposed to be increasingly efficient for these studies. In particular, deep neural networks have been applied to recognize, classify, and characterize complex sets of data. Such methods have proved to be useful and strikingly successful for the investigation of complex physical problems~\cite{MLReview1,MLReview2}.
For example, they have been used in the study of the QCD equation of state by classifying phase transition types~\cite{MLPH1,MLPH2,MLPH3,MLPH4,MLPH5,MLPH6}, in the search for the order parameter of nuclear liquid-gas phase transitions~\cite{MLLGPH1,MLLGPH2}, and in identifying phase transitions by circumventing the fermion sign problem~\cite{MLFermionSign}. One of the most promising advantages of deep learning is that it only requires raw low-level data, such as spin configurations in the Ising model; hence, only elementary knowledge of physics is required. With sufficiently large data sets, higher-level features can be recognized by various ML architectures~\cite{MLArctec}, which can then be used to identify phases.

The last decade has witnessed several exciting explorations into identifying phase transitions in the Ising model  using both  supervised and unsupervised ML techniques~\cite{IsingZ21,IsingZ22,CNNIsing1,MLIsing1,MLIsing2,MLIsing3,MLIsing4,MLIsing5,MLIsing6,MLIsing7,CNNIsing2,CNNIsing4,Feature2DIsing,MLIsing8,CNNIsing3,CriticalandRG1,MLIsing9,MLIsing10}. It has been shown that ML method is able to recognize phases and phase transitions  in various two-dimensional(2D) Ising Hamiltonians~\cite{CNNIsing1,MLIsing1,MLIsing2,MLIsing3,MLIsing4}. The order parameters of multiple types of models can be learned by the weight parameters in the hidden layers of the neural networks~\cite{MLIsing5,MLIsing6,MLIsing7,CNNIsing2,CNNIsing4,Feature2DIsing}. This allows the statistical finite-size scaling analysis to locate the critical temperature and important critical exponents~\cite{MLIsing8,CNNIsing3}. Furthermore, the critical region of the Ising system can be identified, and the symmetry that drives the transition can be reconstructed from the performance of the ML process~\cite{IsingZ22}. However, the majority of current investigations focus on the 2D model, and studies are restricted to only the second-order phase transition of the Ising model. A systematic investigation on the phase diagram and first-order transition of the 3D Ising model remains unexplored using ML methods.

In this study, we explore the ability of a supervised learning method to discriminate phases and classify phase transitions according to configurations produced by a 3D Ising model. We first demonstrate that deep neural networks can learn thermodynamical observables such as magnetization or energy. With a unique network architecture, they can identify both second- and first-order phase transitions in the phase diagram of the Ising model. The important features for the classification of different phases in different types of phase transitions are extracted. The rest of the paper is organized as follows. In Sec. II we give a brief description of the 3D Ising model and its phase structure, explaining how the ML method learns the spin configurations in the model. The results of identifying different types of phase transitions in the Ising model are reported and discussed in Sec. III. Our findings are summarized in Sec. IV.  

\section{Ising Model and Machine Learning Method}
The Ising model~\cite{IsingModel} is regarded as one of the most fundamental magnetic models used to study the nature of phase transitions from a microscopic viewpoint in statistical physics. The energy of any particular state is given by the Ising Hamiltonian: 
\begin{equation}
E=-H\sum_{i}\sigma_i-J\sum_{\langle ij\rangle}\sigma_i\sigma_j.
 \label{Eq:Energy}
\end{equation}
\noindent Here, $\sigma_i$ is the spin of the $i$th site. It equals either 1 (spin up) or -1 (spin down). We consider the model on a cubic lattice with periodic boundary conditions and set $J=1$ as the energy unit. The first sum describes the interaction of the spins with an external magnetic field $H$. The energy is minimum when a spin points parallel to the external magnetic field. The second sum is taken only over pairs $(i,j)$ that are nearest neighbors in the grid, and it describes the interaction of the spins with each other. The interaction energy of a pair of adjacent spins is minimum when they point in the same direction. We usually normalize quantities of interest by the number of degrees of freedom and then work with the average energy per spin, $\epsilon=\langle E\rangle/N$, and the average magnetization per spin, $m=\frac{1}{N}\langle\sum_{i}\sigma_i\rangle$. The average magnetization can be regarded as the order parameter of this model.

\begin{figure}
\hspace{-0.8cm}
\includegraphics[scale=0.36]{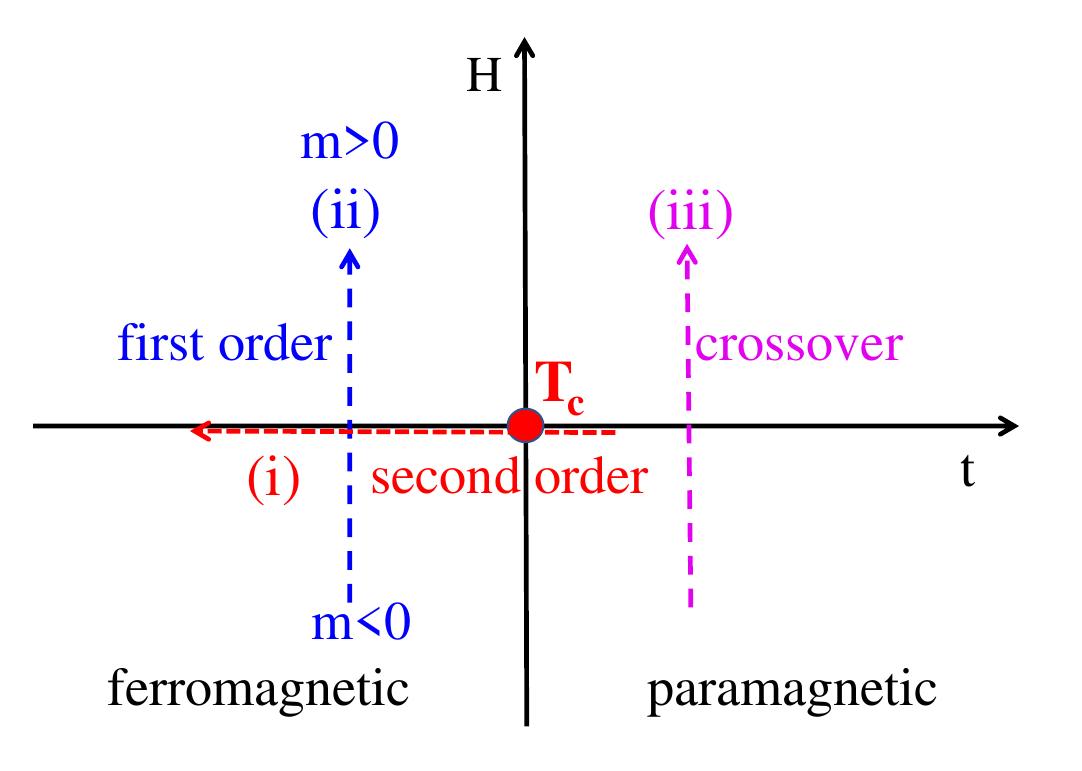}
\caption{(color online) Phase diagram of the 3D Ising model.}
\label{Fig:IsingPhaseD}
\end{figure}

 \begin{figure*}
     \centering
     \includegraphics[scale=0.35]{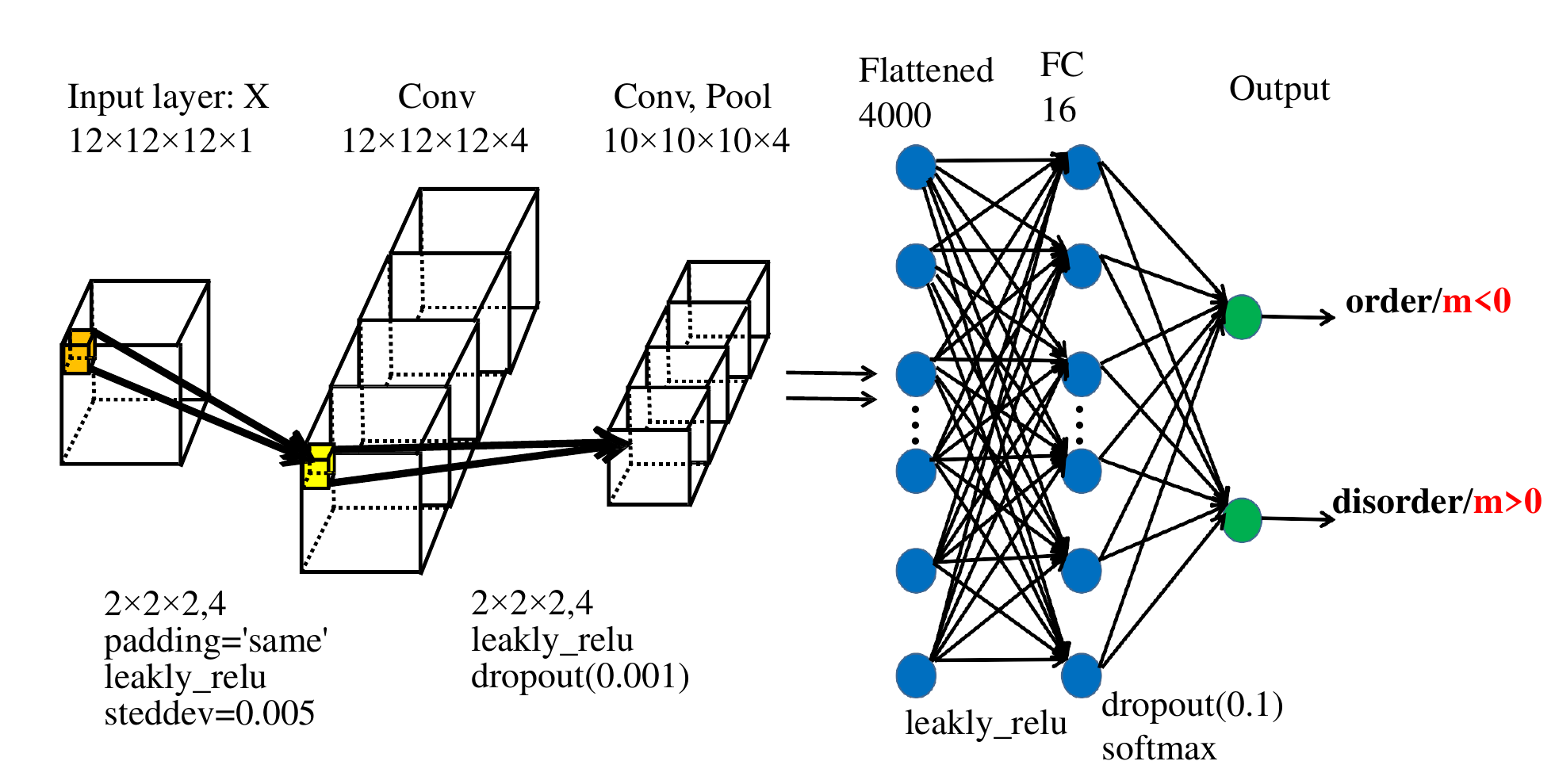}
      \caption{(color online) 3D convolutional neural network architecture used in our analysis.}
     \label{Fig:CNNStructure}
\end{figure*}

The phase diagram of the 3D Ising model~\cite{mapPhaseD2,mapEOS3} is shown in Fig.~\ref{Fig:IsingPhaseD}. The horizontal and vertical axes in the figure are reduced temperature $t=(T-T_c)/T_c$ and external magnetic field $H$, respectively. When decreasing temperature under zero magnetic field, $i.e.$ along the direction of line (i) as shown in the plot, a second-order phase transition with spontaneous twofold symmetry breaking occurs at the CP $T_c$. A discrete Z(2) spin inversion symmetry is broken in the ferromagnetic (order) phase below $T_c$ and is restored in the paramagnetic (disordered) phase at temperatures above $T_c$. On the other hand, when we change the external magnetic field with a fixed temperature below $T_c$, $i.e.$, along the direction of line (ii), a first-order phase transition occurs between two phases corresponding to $m<0$ and $m>0$. The stable phase is the one for which $m$ has the same sign as $H$, but the magnetization remains nonzero even in the limit $H\rightarrow 0$. The system therefore undergoes a first-order phase transition in which $m$ changes discontinuously as it crosses the coexistence curve at $H=0$. The size of the discontinuity decreases with increasing temperature and reaches zero at the critical temperature $T_c$, $i.e.$, the coexistence curve ends at a CP. When we change the external magnetic field with a fixed temperature beyond $T_c$, $i.e.$, along the direction of line (iii), 
the transition is a smooth crossover. 
Based on non-universal mapping between Ising variables $(t,H)$ and QCD coordinates $(T,\mu_B)$ in the phase diagrams, we can construct an equation of state matching the first principle lattice QCD calculations, which can be employed in hydrodynamic simulations of relativistic heavy-ion collision~\cite{mapEOS1,mapEOS2,mapEOS3,mapEOS4}.

A Monte Carlo (MC) simulation of the 3D Ising model on a size $L\times L\times L$ is implemented via the Metropolis algorithm. This algorithm is a simple and widely used approach to generate the canonical ensemble.
It allows us to efficiently obtain a large number of uncorrelated sample configurations over a wide temperature
range. We generate $20000$ independent spin configurations at each selected temperature and external magnetic field. Approximately $90\%$ of event samples are randomly chosen as inputs to train the ML model, and the rest are used to test its performance. To reduce correlations of samples, we take 50 sweeps between every two independent configurations~\cite{sweep}. 

ML as a tool for identifying phase transitions has recently garnered significant attention in this field. Supervised learning, a commonly used method, is performed with a training set to teach models to yield the desired output. The algorithm measures its accuracy through the loss function, which is adjusted until the error is  sufficiently minimized. After training, the neural network can recognize unseen samples and predict the correct label, illustrating that it has learned important features that can be used for classification tasks.

In deep learning, a convolutional neural network (CNN)~\cite{MLArctec} is a class of deep neural network, mostly applied to analyze visual imagery. It is inspired by biological processes in which the connectivity pattern between neurons resembles the organization of the animal visual cortex. The CNN is known as the shift or space invariant of multilayer perceptrons for certain architectures~\cite{shiftinv}. It has the distinguishing features of local connectivity, shared weights, pooling, etc. 
Motivated by previous studies~\cite{CNNIsing1,CNNIsing2,CNNIsing4,CNNIsing3}, we apply ML techniques to the 3D Ising model. Supervised learning with a deep CNN architecture is used to explore the phase transitions and structure of the system.

\section{Results and Discussions}
\subsection{ML the magnetization and energy of spin configuration}
In thermodynamics, a phase transition occurs when there is a singularity in the free energy. In the Ising model, calculations of the average energy of various spin configurations help to obtain free energy. Then, a standard recipe can be followed to find all other quantities of interest. Another important macroscopic quantity that describes the system is the average magnetization (order parameter). Phase transitions often involve the development of some type of order with associated symmetry breaking. The broken symmetry is described by an order parameter that usually increases continuously as the system moves deeper into the ordered phase and measures the degree of order as the phase transition proceeds. 

A detail schematic of the CNN architecture used in this analysis is shown in Fig.~\ref{Fig:CNNStructure}. The network is composed of a 4D input layer of $L\times L\times L\times 1$, two 3D convolutional layers, a pooling layer, a fully connected hidden layer, and an output layer. In addition, the pooling  and fully connected hidden layers are followed by a dropout layer with a rate of 0.001 and 0.1, respectively. In the first convolutional layer, there are four filters with a size of $2\times 2\times 2$ and stride $s = 1$, which are applied on the input configurations and create activations. These activations are further convoluted in the second convolutional layer, which  has the same number, size, and stride of filters as  the first convolutional layer. The second convolutional layer is then forwarded by a pooling layer of size $2\times 2\times 2$ and stride $s=1$. The fully connected layer has 16 neurons, and the output layer is another fully connected layer with softmax activation and two neurons to indicate the result of classification. In this study, all the layers are activated by LeaklyReLU functions, except for the output layer. The mean-square error function is used as a loss function, and the Adam optimizer is utilized to update the weights and biases. 

We perform  supervised learning for spin configurations in the 3D Ising model. The input data for ML networks are obtained from MC sampling at different magnetic fields ranging from -0.9 to 0.9 at $T=3.6$ in a cubic lattice $L = 20$. The spin configurations are labeled by their corresponding average magnetization or energy. The data sets are randomly divided into  training and testing sets. The supervised ML networks are trained in the training sets and predicted in the testing sets. We adopt a 3D CNN architecture that is based on Tensorflow 2.0 and implemented with the Keras library.

\begin{figure}
\hspace{-0.8cm}
\includegraphics[scale=0.4]{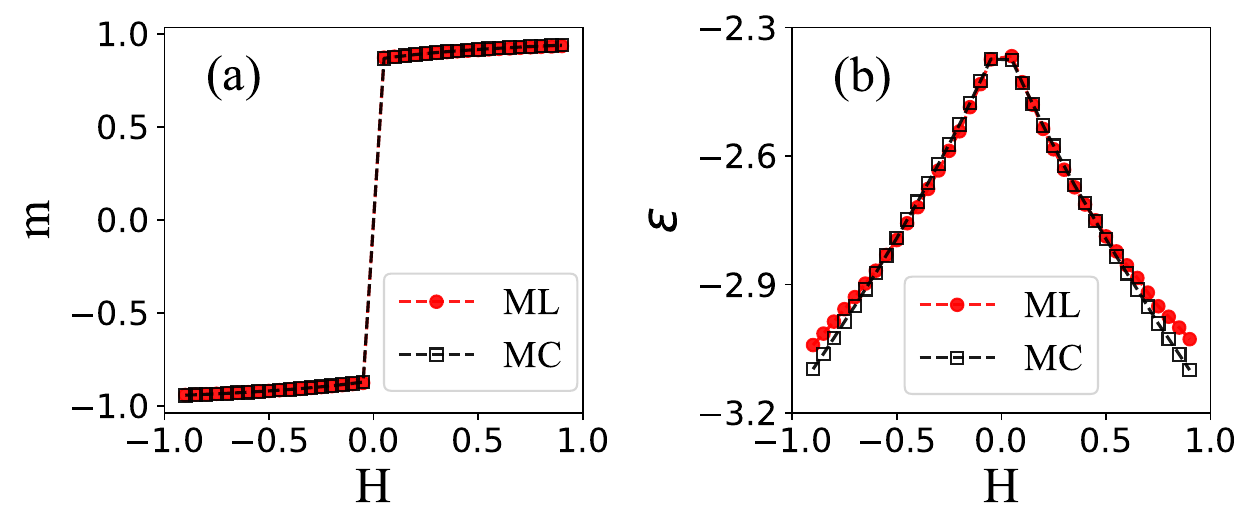}
\caption{(color online) (a) Average magnetization and (b) average energy obtained by predictions of the 3D CNNs (red circles) and Monte Carlo simulations (black squares).}
\label{Fig:MandE}
\end{figure} 

We train the 3D CNNs to predict magnetization and energy of spin configurations and then take an average over the test sets. The results are shown in Fig.~\ref{Fig:MandE} (a) and (b) for average magnetization and average energy, respectively. The red circles represent the averaged predictions from the 3D CNNs, and the black squares are the calculations from MC simulations. The averaged magnetization and energy from predictions of the 3D CNNs agree  well with those from MC simulations. This means that the machine can learn physical quantities of order parameter and energy from pure spin configurations in the Ising model.
\subsection{ML the second-order phase transition}
 \begin{figure*}
     \centering
     \includegraphics[scale=0.5]{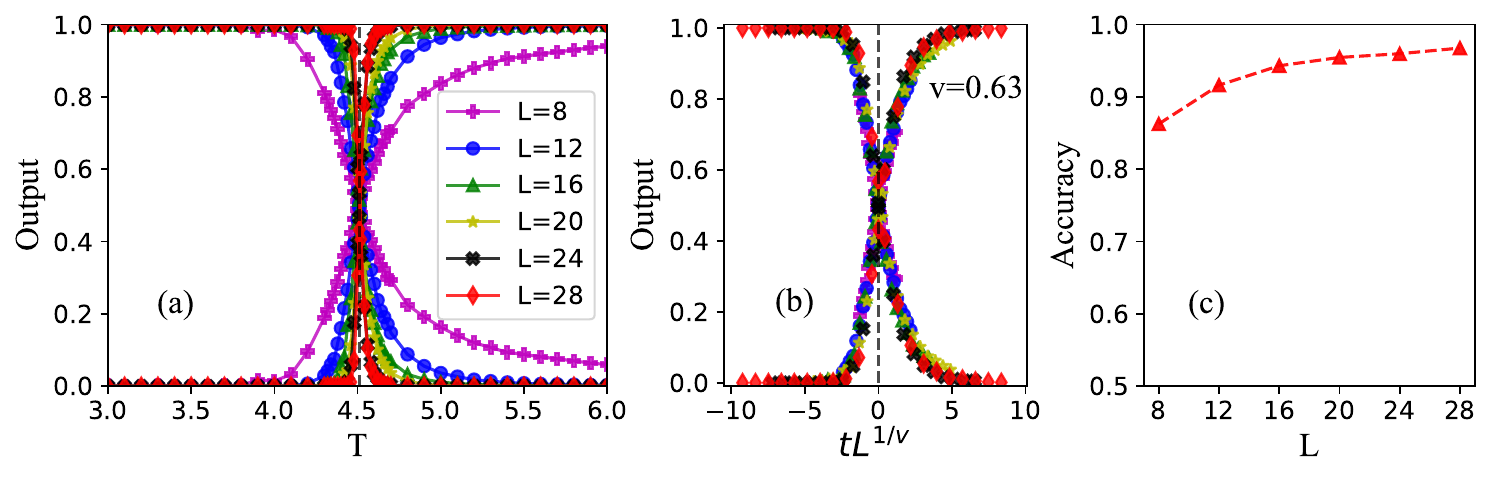}
      \caption{(color online) (a) Output layer averaged over test sets as a function of temperature for six different system sizes. (b) Data collapse of the average output layer as a function of $tL^{\frac{1}{\nu}}$, where $t=(T-T_c)/T_c$ is the reduced temperature. (c) Overall accuracies as a function of $L$. }
     \label{Fig:2ndPH}
\end{figure*}

We now consider the ability of ML to identify phase transitions in the 3D Ising model. As described in Sec. II, the second-order phase transition of the 3D Ising model occurs at a critical temperature $T_c$. It separates a ferromagnetic phase, characterized by a nonzero average magnetization per spin, from a featureless paramagnetic phase at high temperatures. For a given cubic size lattice, each sample in the MC simulations is binary labeled with its corresponding phase, $i.e.$, the high-temperature disorder phase $(0, 1)$ when $T > T_c$ and the low-temperature order phase $(1, 0)$ when $T < T_c$. This gives us the opportunity to attempt to classify the two different types of configurations without use of any thermodynamical estimators. The problem of identifying the phase for a certain value of the temperature is then reformulated as a classification problem in ML.

We construct a 3D CNN to perform supervised learning directly on the uncorrelated raw configurations sampled using MC. The numerical results obtained at various system sizes with $T\in (3.0,6.0)$ at a vanishing magnetic field are illustrated in Fig.~\ref{Fig:2ndPH} (a). The average outputs of different sizes cross at a certain temperature close to that obtained  from Monte Carlo simulations~\cite{3DIsingCriMC1,3DIsingCriMC2}. The CNN model successfully classifies the high- and low-temperature phases. The result is solely based on the raw spin configurations without providing any other thermodynamical quantities. The overall accuracies obtained from the test sets with different system sizes $L$ are shown in Fig.~\ref{Fig:2ndPH} (c). The prediction accuracy increases from $86.3\%$ (for $L=8$) to $96.7\%$ (for $L=28$).

Critical exponents of the 3D Ising model are important physical quantities to describe the critical behavior of phase transitions. They are universal,$i.e.$,they do not depend on the details of the physical system. If the CP of QCD exists, it may well be that its actual critical exponents are those of the 3D Ising model. The correlation length $\xi$ in an infinite system near CP is expected to diverge as $\xi\sim|T-T_c|^{-\nu}$, where $\nu$ is the critical exponent of correlation length. For a finite system, one expects that the correlation length is proportionate to the system size, and thus $|T-T_c|\sim L^{-1/\nu}$. We perform a finite-size scaling analysis~\cite{finiteSizeS} to extract the critical temperature $T_c$ and the critical exponent $\nu$  using the outputs of different sizes in $L\in[8,28]$. The results are shown in Fig.~\ref{Fig:2ndPH} (b). The data points of various sizes collapse with each other  at $T_c = 4.517\pm 0.003$ and $\nu = 0.63 \pm 0.09$, which agrees well with the results of $T_c = 4.512\pm 0.001$ and $\nu = 0.63 \pm 0.01$ obtained using the MC method~\cite{3DIsingCriMC1,3DIsingCriMC3} and renormalization group theory~\cite{3DIsingCriRGT}. Therefore, ML can not only detect phases and locate the transition temperature $T_c$, but also obtain the correct critical exponent of the 3D Ising model.

 \begin{figure*}
     \centering
     \includegraphics[scale=0.4]{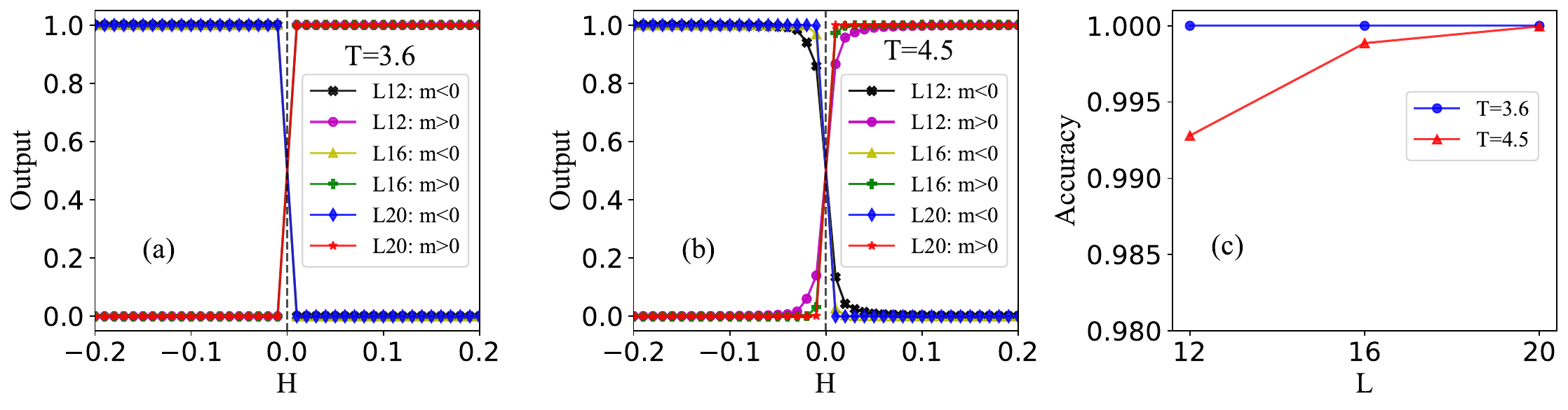}
      \caption{(color online) Output layer averaged over test sets as a function of external magnetic field for three different system sizes at (a) T = 3.6 and (b) T = 4.5. (c) The overall accuracies as a function of $L$ at two different temperatures.}
     \label{Fig:1ndPH}
\end{figure*}
\subsection{ML the first-order phase transition}
The power of neural networks lies in 
their ability to generalize  to  tasks
beyond  their original  design. In the previous subsection, we construct a 
3D CNN to successfully identify the
second-order phase transition in the 3D Ising model. We want to know if the
same network architecture can  also be  used  to  discriminate the  first-order transition in the model.

From the phase diagram of the Ising  model demonstrated in Fig.~\ref{Fig:IsingPhaseD}, the  first-order  phase  transition occurs when we change the external magnetic
field $H$ across the transition boundary $H=0$ with $T<T_c$. To  study  the effect of temperature on this 
transition, we generate event samples by  taking  two  different temperatures of $T=3.6$ (far from the critical temperature $T_c$) and $T=4.5$ (near $T_c$). The same 3D CNN architecture as illustrated in  Fig.~\ref{Fig:CNNStructure} is used to train and make predictions on the test sets for both  cases within the range $H\in [-0.2, 0.2]$. The results are shown in Fig.~\ref{Fig:1ndPH} (a) and (b), respectively. The CNN architecture, which is originally constructed to identify the second-order phase transition, can also successfully classify the $m<0$ and $m>0$ phases. The average outputs of three different sizes cross exactly at $H=0$ in both cases. The test accuracies with different system sizes are shown in Fig.~\ref{Fig:1ndPH} (c). When the temperature of the system is far from $T_c$, the network can classify phases perfectly (blue dots). The accuracy decreases with decreasing system size when the temperature is near $T_c$ (red triangles). This is due to the effect of large critical fluctuations when the system is near the temperature where the second-order phase transition occurs.

\subsection{Exploration of important features in the CNN}

In statistical physics, the second-order  phase transition is  a continuous transition, whereas  the first-order phase  transition  is  a  discontinuous  one. They belong to different types of phase transitions and have different equations of state. The great advantage of the DL method
is the ability to extract hidden features automatically from
a  dataset without the need of human intervention. The neural  network  should  itself select  appropriate  features within the data that are most sensitive to the properties of
different  types  of  equations  of  state.  To  obtain  physical insights  into  how  the  neural-network  discriminates  the phase transitions in the Ising model, it is useful to visualize the complex dependence learned by the network. DL uses multiple layers of representation to learn descriptive features directly  from training  data.  The  important   features,  which  are  used  in  the  classification process  for phase transitions and for identifying any underlying physical quantities, may be encoded in convolutional filter kernels~\cite{FeatureWeightFilter}, weight matrices~\cite{CNNIsing2,FeatureWeightFilter,Feature2DIsing}, or the activations of hidden units in the network~\cite{FeatureActivation}. The sets of variational parameters, specifically the weights and biases at
each layer of the neural network, are optimized during the training process.  They  will  converge  to  certain  values, which encode  the  solution  to  the  particular  problem   under consideration.

 \begin{figure*}
     \centering
     \includegraphics[scale=0.4]{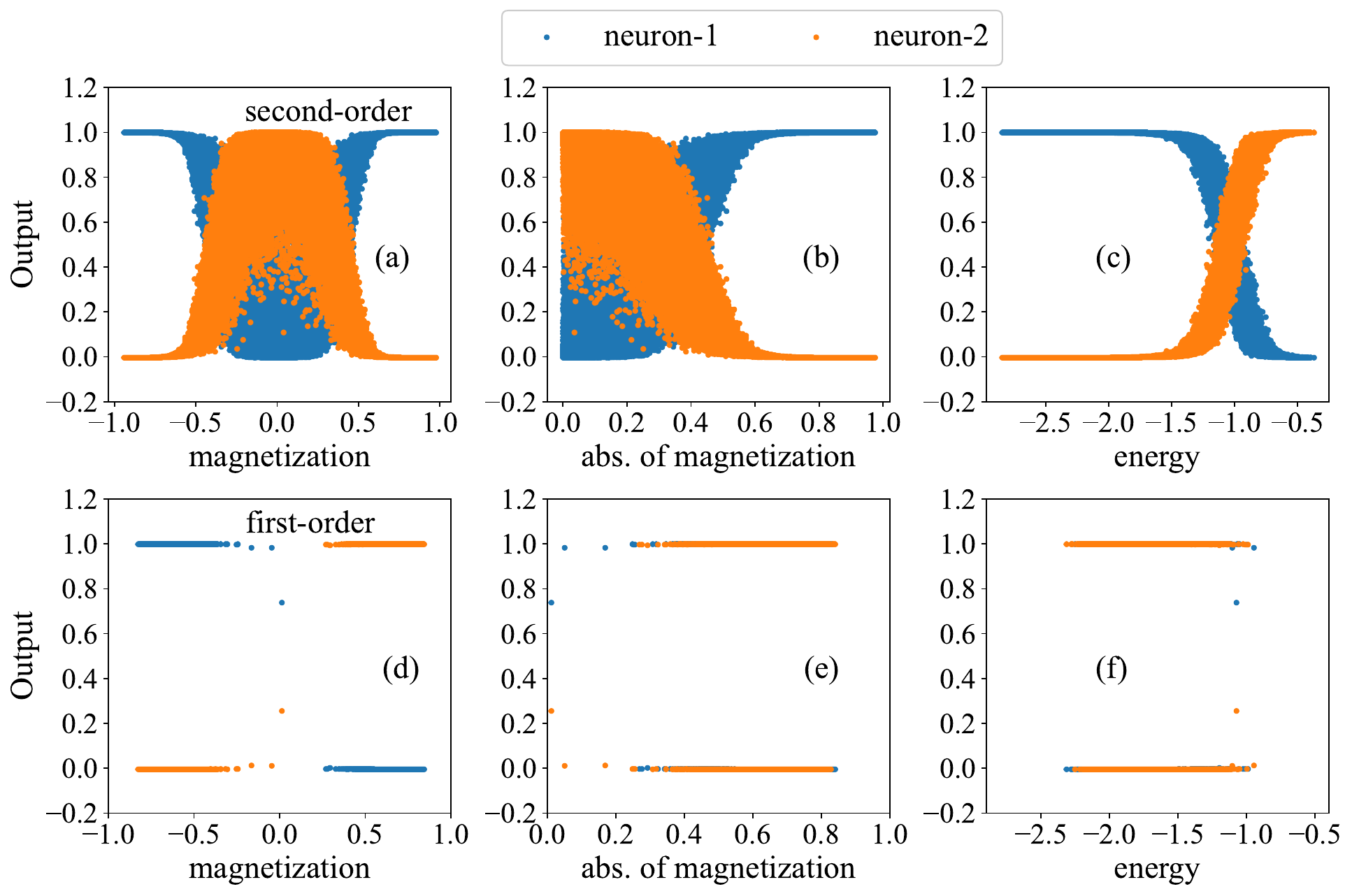}
      \caption{(color online) Correlations between the outputs of two neurons in the output layer of the neural network and magnetization (a)(d), the absolute value of magnetization (b)(e), and energy(c)(f). The first row represents the result from the second-order phase transition, and the second row is that from the first-order phase transition.}
     \label{Fig:corr}
\end{figure*}

We observe that a uniform network  architecture  can identify  different  phases  in  either  the  second-order  or  first-order  phase  transitions  in  the  3D  Ising  model.  It  is
interesting  to  find  out  which  quantities  can  possibly  be used for the classification processes in the results ofFig.~\ref{Fig:2ndPH} and \ref{Fig:1ndPH}. Figure~\ref{Fig:corr} shows a scatter plot of the outputs for two  neurons  in  the  final  output  layer  of  the  CNN  with magnetization,  the absolute  value  of  magnetization, and energy.
From  the  first  row  for  the  second-order  phase transition, we find that the outputs of the network are proportional to the absolute value of magnetization and  energy rather than the magnetization of the input configurations. The calculated absolute values of the Pearson correlation coefficients are 0.17, 0.93, and 0.85 for Fig.~\ref{Fig:corr} (a), (b) and (c), respectively. As for the second row for the first-order transition, the outputs of the network are found to be proportional to magnetization rather than the absolute value of magnetization or energy. The absolute values of the Pearson correlation coefficients are 1, 0.1, and 0.08 for Fig.~\ref{Fig:corr} (d), (e) and (f), respectively.

Within the shallow layers of a network architecture, the variational parameters  correspond  to  learned universal features. This form of universality  is  diminished   toward  deeper layers, where  the features  are  expected  to transit from  universal  to  specific  in relation  to  different systems. The activation function of a variable in an intermediate layer of the neural network acts as a transformation that maps a certain input to an output representation\cite{FeatureActivation}. 

 \begin{figure*}   
    \centering
    \includegraphics[scale=0.3]{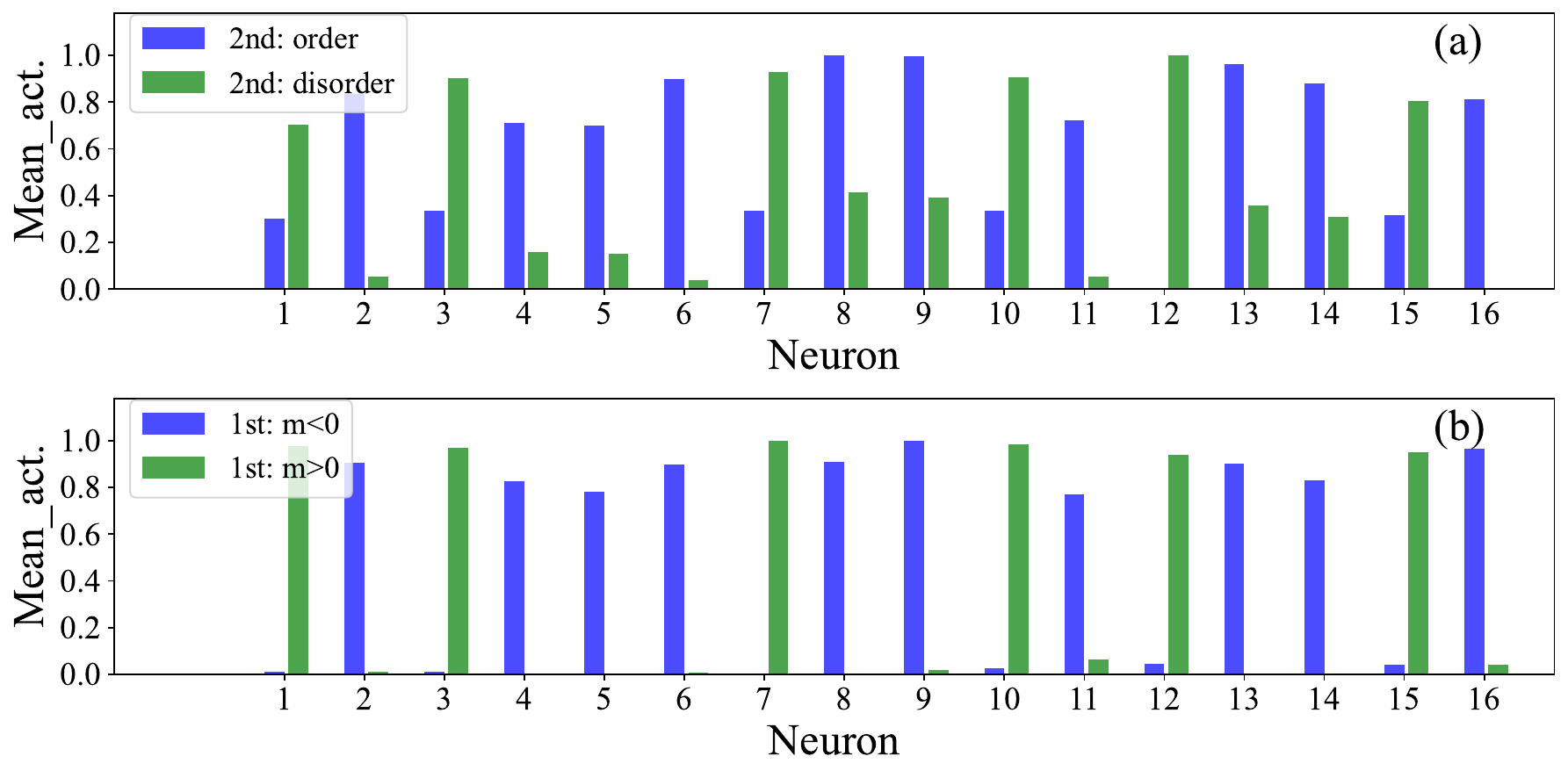}
    \caption{(color online) Mean activations of the 16 hidden neurons in the fully connected layer of the 3D Ising-trained convolutional neural network of different phases in the second-order (a) and the first-order (b) phase transitions.}
    \label{Fig:feature}
    
 \end{figure*} 

We calculate the mean activation functions of neural
network configurations of different phases with $L = 12$ in the second- and first-order phase transitions. The results are shown in Fig.\ref{Fig:feature}, where the activations are drawn for
16  neurons  of  the  fully  connected  layer.  The  blue  and
green  bars  represent  the  calculations  from  different
phases  in  the  same  transitions.  The  maximum  value  is
normalized to  one,  and  the  results  are  rescaled  accordingly for better representation. In Fig.~\ref{Fig:feature} (a), we observe that  nearly  all  of  the  16  neurons  for  both  the  order  and disorder  phases  are  activated  in  the  second-order  phase transition, although the values of the two phases are different from each other. This may be due to large fluctuations  near  the  CP  in  this  transition.  This  might  also  be
reminiscent of the connection between DL and the renormalization  group
\cite{CriticalandRG1,CriticalandRG2}. As for the first-order phase transition in Fig.~\ref{Fig:feature} (b), it is found that different phases ($m<0$ and $m>0$) activate different neurons.
\section{Conclusions and Outlook}
In this paper, we employ a supervised learning method to study phase transitions in the 3D Ising model. It is found that a 3D CNN can be trained to predict the magnetization and  energy  of  spin  configurations.  The   network successfully classifies the high- and low-temperature  phases  of  the  3D  Ising  model. It  is  able  to  not  only locate the transition temperature but also obtain the critical exponent of the second-order phase transition based on  raw spin configurations using the finite-size scaling analysis method. With the same CNN architecture, the neural network can identify different phases in the first-order transition with  high  accuracy. This  implies  that  the   network  captures  both  the  second-  and  first-order  phase transitions and encodes them in the parameters of the network during the process of supervised learning. Furthermore, we  attempt to extract and decode important features of the classification processes directly from the data. It is found that the neural network detects phases by the absolute  value  of  magnetization  and  energy  in  the second-order phase transition, whereas it tends to rely on magnetization in the first-order transition. The calculated mean  activations  in  the  high-level  representations  show that almost all the neurons before the output layer are activated in the second-order phase transition, but different
phases activate different neurons in the case of the first-order.

In relativistic heavy-ion collisions, we expect the QCD CP to have similar critical behaviors to those in the 3D Ising model owing to the same universality that is dictated by  the  symmetry  of  the  systems.  In  heavy-ion experiments, the measured  high-order  cumulants
\cite{net_proton2014,STARcumulant,STARPRCMoment} and intermittency~\cite{NA49SFM,NA61universe,STARintermittency,overview} both at SPS and RHIC energies suggest that there are large density  fluctuations near the QCD CP. Here, we prove that the state-of-the-art ML method can learn important features  that discriminate different  phases  in the phase transition. Understanding what this approach captures will be useful when we experimentally explore the QCD CP and phase boundary. The method developed in this study can also be applied to investigate the criticality in MC models, which could help to  explore  the  physical  mechanisms  of  the  QCD  phase transition.

\section*{Acknowledgments}
We are grateful to Prof. Longgang Pang and Mingmei Xu for their fruitful discussions and comments. We further thank Prof. Hengtong Ding for providing us with computing resources. The numerical simulations have been performed on the GPU cluster in the Nuclear Science Computing Center at Central China Normal University ($\rm{NSC^3}$). 
This work is supported by the National Natural Science Foundation of China (No. 12275102) and the National Key Research and Development Program of China (No. 2022YFA1604900). 

X.L. and R.G. contributed equally to this work.


\end{document}